\documentclass[lettersize,journal]{IEEEtran}
\usepackage{amsmath,amsfonts}
\usepackage{algorithm}
\usepackage{array}
\usepackage[caption=false,font=normalsize,labelfont=sf,textfont=sf]{subfig}
\usepackage{textcomp}
\usepackage{stfloats}
\usepackage{url}
\usepackage{verbatim}
\usepackage{graphicx}
\usepackage{cite}
\usepackage{algpseudocode}
\usepackage{booktabs} 
\usepackage{xcolor} 

\begin{document}


\title{A Dynamic Recurrent Adjacency Memory Network for Mixed-Generation Power System Stability Forecasting}



\author{
Guang~An~Ooi, Otavio~Bertozzi, Mohd~Asim~Aftab,~\IEEEmembership{Senior~Member,~IEEE},
Charalambos~Konstantinou,~\IEEEmembership{Senior~Member,~IEEE},~
and Shehab~Ahmed,~\IEEEmembership{Senior~Member,~IEEE}%
\thanks{G. A. Ooi, O. Bertozzi,  M. A. Aftab, C. Konstantinou, and S. Ahmed are with King Abdullah University of Science and Technology (KAUST), Thuwal~23955, Saudi~Arabia.}
}

\markboth{}%
{Ooi \MakeLowercase{\textit{et al.}}: A Dynamic Recurrent Adjacency Memory Network for Mixed-Generation Power System Stability Forecasting}


\maketitle

\begin{abstract}
Modern power systems with high penetration of inverter-based resources exhibit complex dynamic behaviors that challenge the scalability and generalizability of traditional stability assessment methods. This paper presents a dynamic recurrent adjacency memory network (DRAMN) that combines physics-informed analysis with deep learning for real-time power system stability forecasting. The framework employs sliding-window dynamic mode decomposition to construct time-varying, multi-layer adjacency matrices from phasor measurement unit and sensor data to capture system dynamics such as modal participation factors, coupling strengths, phase relationships, and spectral energy distributions. As opposed to processing spatial and temporal dependencies separately, DRAMN integrates graph convolution operations directly within recurrent gating mechanisms, enabling simultaneous modeling of evolving dynamics and temporal dependencies. Extensive validations on modified IEEE 9-bus, 39-bus, and a multi-terminal HVDC network demonstrate high performance, achieving 99.85\%, 99.90\%, and 99.69\% average accuracies, respectively, surpassing all tested benchmarks, including classical machine learning algorithms and recent graph-based models. The framework identifies optimal combinations of measurements that reduce feature dimensionality by 82\% without performance degradation. Correlation analysis between dominant measurements for small-signal and transient stability events validates generalizability across different stability phenomena. DRAMN achieves state-of-the-art accuracy while providing enhanced interpretability for power system operators, making it suitable for real-time deployment in modern control centers.
\end{abstract}


\section{Introduction}
\IEEEPARstart{T}{he} assessment and forecasting of stability in modern power systems, characterized by high penetration levels of inverter-based resources (IBRs), presents growing challenges~\cite{Hatziargyriou2021,alimi2020review,Milano2018}. Traditional stability assessment methods, such as eigenvalue analysis for small-signal conditions and time-domain simulations for transient behavior, are based on offline studies of fixed system models. These methods, though reliable under conventional setups, encounter limitations in scalability and accuracy as the system operating point evolves more rapidly due to the variability introduced by IBRs~\cite{zhou2024efficient}. As the share of IBRs increases, the dynamic interactions between synchronous generators (SGs), grid-forming (GFM), and grid-following (GFL) converters become more complex, causing significant variation in stability margins and transient response characteristics. These evolving dynamics challenge the assumptions underlying classical techniques, motivating the exploration of data-driven methods as a promising alternative. Such approaches aim to deliver fast and generalizable stability forecasts using real-time measurements, thereby reducing reliance on computationally intensive simulations and fixed system models~\cite{bertozzi2024data,alimi2020review}.
\vspace{-2mm}
\subsection{Related Work}
Recent research has explored a variety of strategies for stability assessment, ranging from classical dynamic analysis to modern machine learning (ML) models. Each category offers distinct advantages and limitations depending on the operating context, system observability, and required interpretability. 
\begin{table*}[!t]
    \renewcommand{\arraystretch}{1.3}
    \caption{Selected References on Stability Forecasting and Assessment}\label{tab:stability_refs}
    \centering
    \small
    \begin{tabular}{l c c}
        \hline\hline \\[-3mm]
        Paper & Technique & Application \\[1.4ex] \hline
        ~\cite{kelada2025revisiting,lara2023revisiting}   & Modal / time-domain analysis & Foundational baselines \\
        ~\cite{chen2024real}      & GNNs & Stability risk assessment \\
        ~\cite{zhu2024scaling}      & GNNs & Large-scale analysis \\
        ~\cite{huang2024multi}      & GraphSAGE-A & Multi-task transient stability assessment \\
        ~\cite{zhao2024swin}       & Swin Transformer & Transient stability assessment \\
        \textit{This paper} & DRAMN& Unified real-time small-signal and transient forecasting \\
        \hline\hline
    \end{tabular}
\end{table*}

\subsubsection{Classical Methods for Stability Assessment}
Small-signal and transient stability in power systems are traditionally assessed using eigenvalue-based modal analysis and time-domain simulations. Modal analysis enables the study of system dynamics by linearizing the system around a steady-state operating point~\cite{kelada2025revisiting}. It is effective for tuning power system stabilizers and evaluating oscillatory behavior under small disturbances. However, its reliance on accurate linear models and fixed operating points limits its adaptability to fast-changing system conditions. Time-domain simulations numerically solve differential algebraic equations to capture the full nonlinear response under large disturbances. While highly accurate, especially for transient stability analysis, these approaches are computationally intensive and unsuitable for real-time applications~\cite{lara2023revisiting}. Both fundamental techniques struggle to scale under high variability and limited observability conditions induced by widespread IBR deployment.

\subsubsection{Data-Driven Machine Learning Models}
To overcome limitations in classical stability assessment methods, data-driven techniques have been increasingly adopted for stability classification and prediction. Classical ML models such as support vector machines, decision trees, random forests, and XGBoost have been applied using static features like pre-fault voltages, rotor angles, and bus frequencies~\cite{alimi2021review}. While these methods are lightweight and easy to interpret, they often rely on manual feature engineering and fail to capture temporal dependencies inherent in system dynamics.

Recurrent models such as long short-term memory recurrent neural networks (LSTM-RNNs) and gated recurrent units have shown promise for capturing sequential patterns in phasor measurement unit (PMU) data~\cite{alimi2021review, bertozzi2024data}. These approaches improve temporal awareness but typically treat each system node independently, ignoring the spatial and structural dependencies of the grid. As a result, they may overfit to specific scenarios and exhibit poor generalization under unseen operating conditions.

\subsubsection{Graph Neural Networks in Power Systems}
Graph neural networks (GNNs) provide a natural framework for modeling power systems, where buses and lines form an inherent graph structure. Early GNN applications in this domain use graph convolutional networks (GCNs) or graph attention networks (GATs) to incorporate topological information into load forecasting, fault localization, or stability classification tasks~\cite{chen2024real,zhu2024scaling}. However, most of these models use static adjacency matrices derived from physical bus connectivity or line admittance, which do not capture dynamic electrical interactions or modal coupling effects that vary with generation mix, load level, or fault type.

More recent work has proposed spatial-temporal GNNs to learn from time-varying graph signals, using snapshots of grid measurements as node features. These approaches offer improved forecasting performance but still rely on heuristic or static definitions of graph structure, failing to reflect the true modal dynamics of the system.

\subsubsection{Physics-Informed and Hybrid Approaches}
To address the gap between physical fidelity and data-driven generalization, hybrid models have been developed to incorporate physics-based features or constraints into learning frameworks. Some studies include inertia constants, damping coefficients, or participation factors as input features to neural networks~\cite{linaro2023continuous, cao2023physics}, while others enforce physical laws via regularization terms in the loss function~\cite{falas2025robust}. These models improve interpretability and generalization but often require extensive modeling effort or access to detailed system parameters.

A promising direction involves the use of dynamic mode decomposition (DMD) to extract dominant spatiotemporal modes from measurement data. DMD has been applied for post-event oscillation analysis, inertia estimation, and Koopman operator approximation in power systems~\cite{korda2018convergence}. Variants such as time-delayed DMD (TD-DMD)~\cite{nedzhibov2025blind}, kernel extended DMD (KE-DMD)~\cite{philipp2025error}, and Hankel DMD~\cite{sakib2025learning} improve robustness under noise and nonlinearity. However, most DMD applications remain limited to offline analysis and are not integrated with graph-based learning or used as inputs to real-time forecasting architectures. The framework proposed in this study overcomes these limitations by integrating DMD-derived spectral interactions directly within recurrent graph convolutions, enabling real-time learning of evolving modal relationships. In contrast to previous DMD or GNN methods, it jointly models temporal and spatial dependencies through dynamic adjacency updates, providing unified and interpretable stability forecasting across varying operating conditions.

\subsection{Contributions}
This paper proposes a dynamic recurrent adjacency memory network (DRAMN) for power system stability forecasting. Sliding-window DMD is employed to construct multi-layer adjacency matrices that capture evolving modal interactions, while graph convolutions embedded within recurrent cells enable joint modeling of spatial and temporal dependencies. The framework provides unified prediction of small-signal and transient instability, achieves 82\% feature reduction without loss of accuracy, and is validated on modified IEEE 9-bus and multi-terminal HVDC systems, with scalability demonstrated on the IEEE 39-bus benchmark. The main contributions of this paper are listed as follows:
\begin{itemize}
    \item Integration of sliding-window DMD with recurrent graph convolutions, enabling joint spatio–temporal modeling through physics-informed adjacency updates within LSTM gating mechanisms.
    \item Construction of a multi-layer dynamic adjacency tensor that captures distinct modal properties and supports unified forecasting of small-signal and transient instability across diverse systems.
    \item Achievement of real-time, millisecond-level stability prediction with strong generalization and feature reduction, allowing deployment under limited observability.
\end{itemize}

The remainder of this paper is organized as follows: Section II introduces the DRAMN framework, covering adjacency matrix construction and recurrent graph integration. Section III outlines the simulation setup, dataset generation, and labeling. Section IV reports evaluation results, including sensitivity analysis, benchmarking, ablation, and case studies on HVDC and 39-bus systems. Section V discusses scalability, noise sensitivity, and real-time feasibility. Section VI concludes with key findings and future directions.

\section{DRAMN Stability Forecasting}
\begin{figure*}[t!]
    \centering
    \includegraphics[width=0.8\linewidth]{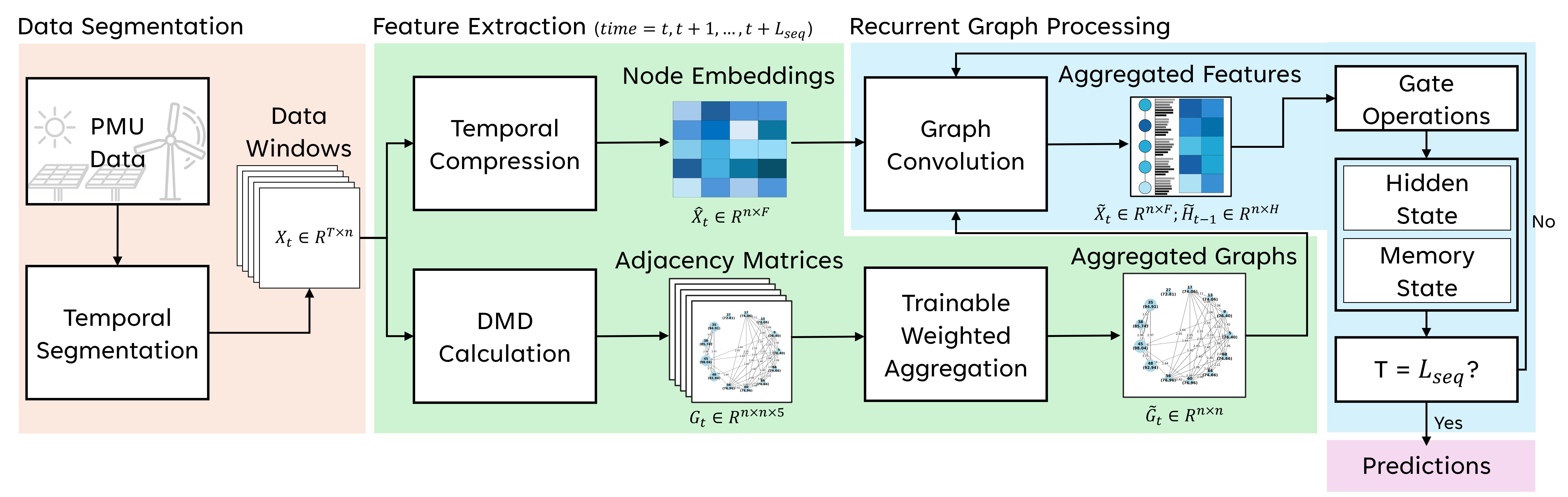}
    \caption{Overview of the proposed DRAMN framework. Time-series measurements are processed in a sliding window to extract modal features using DMD. These features are mapped to multi-layer dynamic adjacency matrices representing evolving modal interactions. The recurrent network utilizes both node features and dynamic graph structures to forecast small-signal and transient instability probabilities.}
    \label{fig:dramn}
\end{figure*}

The proposed framework extracts evolving multi-layer adjacency matrices from time-series data by applying DMD over sliding windows of arbitrary duration to capture inter-component interactions. The temporal compression stage generates compact node embeddings that preserve essential system state information while reducing computational complexity. These multi-dimensional representations are then processed through a recurrent GNN architecture that performs graph convolution operations on the adjacency matrices to extract spatial features, followed by trainable weighted aggregation to combine information across the five adjacency layers. The overall framework is illustrated in Fig.~\ref{fig:dramn}.

\subsection{Sliding-Window DMD for Adjacency Matrix Extraction}
This section outlines the construction of multi-layer adjacency matrices from raw time-series data using a sliding-window DMD framework. DMD is a data-driven technique that approximates system dynamics by decomposing time-series snapshots into spatial modes associated with growth or decay rates and oscillatory frequencies~\cite{korda2018convergence}. In this work, DMD is applied over sliding windows to capture the evolving spectral content of the system, thereby reflecting transient variations in modal interactions. For each window, the decomposition is truncated to rank $r=5$, which retains the dominant electromechanical and converter-driven modes typically governing stability, while filtering out noise and reducing computational complexity. The resulting modes form a five-layer adjacency tensor, with each layer encoding a distinct dynamic property of the grid.

A linear operator $A$ is estimated via DMD to characterize the temporal evolution of discrete-time system states $\mathbf{x}_{k+1} \approx A\,\mathbf{x}_k$, where $A \in \mathbb{R}^{n \times n}$ and $\mathbf{x}_k \in \mathbb{R}^n$ is the state at sample $k$. For a window of $T$ samples with sampling interval $\Delta t$, the snapshot matrices $\mathbf{X}_1$ and $\mathbf{X}_2$ are formed in (\ref{eq:X1})-(\ref{eq:X2}):
\begin{equation}
\mathbf{X}_1 = \begin{bmatrix}\mathbf{x}_1 & \mathbf{x}_2 & \cdots & \mathbf{x}_{T-1}\end{bmatrix} \in \mathbb{R}^{n \times (T-1)},
\label{eq:X1}
\end{equation}
\begin{equation}
\mathbf{X}_2 = \begin{bmatrix}\mathbf{x}_2 & \mathbf{x}_3 & \cdots & \mathbf{x}_{T}\end{bmatrix} \in \mathbb{R}^{n \times (T-1)}.
\label{eq:X2}
\end{equation}
Computing the thin singular value decomposition of $\mathbf{X}_1 = \mathbf{U}\boldsymbol{\Sigma}\mathbf{V}^\top$ and truncating to rank $r = 5$ provides  $\mathbf{U}_r,\boldsymbol{\Sigma}_r,\mathbf{V}_r$. The reduced operator and its eigendecomposition are shown in (\ref{eq:reduced_operator_A})-(\ref{eq:eigen_decomp_operator_A}):
\begin{equation}
\tilde{A} = \mathbf{U}_r^\top\,\mathbf{X}_2\,\mathbf{V}_r\,\boldsymbol{\Sigma}_r^{-1},
\label{eq:reduced_operator_A}
\end{equation}
\begin{equation}
\tilde{A}\,\mathbf{W} = \mathbf{W}\,\boldsymbol{\Lambda}.
\label{eq:eigen_decomp_operator_A}
\end{equation}
The columns of $\mathbf{W}$ are eigenvectors and $\boldsymbol{\Lambda}=\mathrm{diag}(\lambda_1,\ldots,\lambda_r)$ contains the DMD eigenvalues. (\ref{eq:dmd_phi}) calculates the full-state DMD modes:
\begin{equation}
\boldsymbol{\Phi} = \mathbf{X}_2\,\mathbf{V}_r\,\boldsymbol{\Sigma}_r^{-1}\,\mathbf{W}.
\label{eq:dmd_phi}
\end{equation}
These eigenvalues characterize the growth and decay rates as well as frequencies of the dynamic modes. Each column of $\mathbf{\Phi}$ represents a coherent spatio-temporal pattern in the original data, and its corresponding eigenvalue in $\mathbf{\Lambda}$ dictates its temporal evolution. Here, $\mathbf{\Phi}\in\mathbb{C}^{n\times r}$ collects the retained DMD modes, and its entries are denoted by $\phi_{ik}$.

\subsection{Sequential Multi-layer Adjacency Construction} 
For the DRAMN framework, an arbitrary sequence of $L_{\text{seq}} = 5$ consecutive time windows is processed through DMD to construct a temporal sequence of multi-layer adjacency matrices. For each time window $t$, an adjacency matrix $\mathcal{G}_t \in \mathbb{R}^{n \times n \times d}$ is constructed, where $d = 5$ represents the number of spectral layers. Each layer in $\mathcal{G}_t$ encodes a distinct physical interpretation of dynamic interdependencies. The first layer $\mathcal{G}_t(:,:,1)$ encodes element-wise participation factors, indicating the mutual involvement of elements $i$ and $j$ across dynamic modes. Each element is computed based on the absolute values of the DMD mode entries, as shown in (\ref{eq:Gt1}):
\begin{equation}
\mathcal{G}_t(i,j,1) = \sum_{k=1}^{r} |\phi_{ik}| \cdot |\phi_{jk}|.
\label{eq:Gt1}
\end{equation}
The second layer $\mathcal{G}_t(:,:,2)$ captures coupling strength by utilizing the average mode magnitudes associated with each system element, as shown in (\ref{eq:Gt2}). Defining the per-node magnitude vector as $\mathbf{v}_i = \big[\,|\phi_{i1}|,\ldots,|\phi_{ir}|\,\big]^\top$, its mean $\bar v_i = \frac{1}{r}\sum_{k=1}^r |\phi_{ik}|$, and the zero-mean vector $\tilde{\mathbf{v}}_i=\mathbf{v}_i-\bar v_i\mathbf{1}$, where $\mathbf 1\in\mathbb{R}^{r}$ denotes the all-ones vector, each element in the second layer can be calculated as in (\ref{eq:Gt2})
\begin{equation}
\mathcal{G}_t(i,j,2) =
\dfrac{\tilde{\mathbf{v}}_i^\top \tilde{\mathbf{v}}_j}
      {(\|\tilde{\mathbf{v}}_i\|_2 + \varepsilon)(\|\tilde{\mathbf{v}}_j\|_2 + \varepsilon)},
\quad \mathcal{G}_t(i,j,2) \in [-1,1],
\label{eq:Gt2}
\end{equation}
where $\varepsilon = 10^{-8}$ to avoid zero divisions.
The third layer encodes phase similarity. (\ref{eq:theta_components}) computes the circular mean components for each node $i$:
\begin{equation}
c_i=\frac{1}{r}\sum_{k=1}^r \cos\!\big(\angle\phi_{ik}\big),\qquad 
s_i=\frac{1}{r}\sum_{k=1}^r \sin\!\big(\angle\phi_{ik}\big),
\label{eq:theta_components}
\end{equation}
and define
\begin{equation}
\theta_i=\operatorname{atan2}(s_i,c_i),\qquad \kappa_i=\sqrt{c_i^2+s_i^2}\in[0,1] .
\label{eq:theta_i}
\end{equation}
A bounded, shift-invariant phase similarity for each $i$ and $j$ is then defined in (\ref{eq:Gt3})
\begin{equation}
\mathcal{G}_t(i,j,3) = \kappa_i\,\kappa_j\,\cos\!\big(\theta_i-\theta_j\big) .
\label{eq:Gt3}
\end{equation}
The fourth layer represents co-activation patterns weighted by per-step growth/decay. Let $\lambda_k$ be the DMD eigenvalues and $\rho_k = |\lambda_k|$. Defining a signed, dimensionless growth measure as $g_k = \log \rho_k$, and $\mathbf{D}_{\mathrm{grow}}^{(d)}=\mathrm{diag}(g_1,\ldots,g_r)$, the spectral projection is calculated in (\ref{eq:Gt4}):
\begin{equation}
\mathcal{G}_t(:,:,4) =  \operatorname{Re}\!\big(\boldsymbol{\Phi}\,\mathbf{D}_{\mathrm{grow}}^{(d)}\,\boldsymbol{\Phi}^{\mathrm H}\big) .
\label{eq:Gt4}
\end{equation}
The fifth layer encodes windowed spectral energy. Let $L$ be the number of samples in the window. For each mode $k$, the discrete energy factor $e_k(L)$ is computed as in \eqref{eq:eng_weight}:
\begin{equation}
e_k(L)=
\begin{cases}
  \dfrac{\operatorname{expm1}(2L\log\rho_k)}{\operatorname{expm1}(2\log\rho_k)}, & \rho_k\neq 1,\\[6pt]
  L, & \rho_k=1~,
\end{cases}
\qquad \rho_k=|\lambda_k| .
\label{eq:eng_weight}
\end{equation}
With $\mathbf{D}_{\mathrm{eng}}^{(d)}=\mathrm{diag}\!\big(e_1(L),\ldots,e_r(L)\big)$, the resulting projection is given in \eqref{eq:Gt5}:
\begin{equation}
\mathcal{G}_t(:,:,5) =  \operatorname{Re}\!\big(\boldsymbol{\Phi}\,\mathbf{D}_{\mathrm{eng}}^{(d)}\,\boldsymbol{\Phi}^{\mathrm H}\big) .
\label{eq:Gt5}
\end{equation}
This weights modes by their total energy over the $L$‑sample window, naturally favoring persistent or amplifying modes and de‑emphasizing strongly decaying ones.

Following their construction, each of the $d$ layers is normalized by its maximum value to ensure uniform scaling and make $\mathcal{G}_t$ suitable for spatio-temporal learning. The sequence ${\mathcal{G}_1, \mathcal{G}_2, \mathcal{G}_3, \mathcal{G}_4, \mathcal{G}_5}$ captures the temporal evolution of spatio-spectral relationships across five consecutive time windows.

\subsection{The DRAMN Architecture}
The proposed DRAMN possesses a recurrent neural architecture as shown in Fig.~\ref{fig:dramn}. The input to the network comprises a sequence of raw measurements $X \in \mathbb{R}^{L_{\text{seq}} \times T \times n}$ and the corresponding sequence of multi-layer adjacency matrices $G \in \mathbb{R}^{L_{\text{seq}} \times n \times n \times d}$, where $L_{\text{seq}} = 5$ denotes the sequence length (number of consecutive time steps), $T$ denotes the local temporal window size for each time step, $n$ the number of system elements (nodes), and $d = 5$ the number of spectral adjacency layers. This sequential structure enables DRAMN to capture both the evolution of spatial relationships through the changing adjacency matrices and temporal dependencies through recurrent processing of the five-step sequence.

At each time step $t$, the slice $X_t \in \mathbb{R}^{T \times n}$ is temporally compressed via a $1 \times 1$ convolution, followed by global average pooling and a fully connected projection. This produces a latent node-wise embedding $\hat{X}_t \in \mathbb{R}^{n \times F}$, where $F = 64$ denotes the embedding dimension.

Concurrently, the multi-layer adjacency tensor $\mathcal{G}_t \in \mathbb{R}^{n \times n \times d}$ is aggregated into an effective adjacency matrix using a trainable weighted combination $\tilde{G}_t = \sum_{k=1}^{d} \alpha_k \mathcal{G}_t(:,:,k)$, where $\alpha_k \in \mathbb{R}$ are learnable scalars for each spectral layer $k$.

The input and hidden states are then graph-convolved using $\tilde{G}_t$ as shown in (\ref{eq:Xt})-(\ref{eq:Ht-1}):
\begin{equation} 
\tilde{X}_t = \tilde{G}_t \hat{X}_t 
\label{eq:Xt}
\end{equation}
\begin{equation} 
\tilde{H}_{t-1} = \tilde{G}_t h_{t-1}.
\label{eq:Ht-1}
\end{equation}
Gate pre-activations are obtained by applying graph convolutions to both the input embeddings and prior hidden states using $\tilde{G}_t$. Each gate is computed separately as shown in (\ref{eq:it})-(\ref{eq:gt}):
\begin{equation}
i_t = \sigma\left( \tilde{X}_t W_{x_i} + \tilde{H}_{t-1} W_{h_i} + b_i \right),
\label{eq:it}
\end{equation}
\begin{equation}
f_t = \sigma\left( \tilde{X}_t W_{x_f} + \tilde{H}_{t-1} W_{h_f} + b_f \right),
\label{eq:ft}
\end{equation}
\begin{equation}
o_t = \sigma\left( \tilde{X}_t W_{x_o} + \tilde{H}_{t-1} W_{h_o} + b_o \right),
\label{eq:ot}
\end{equation}
\begin{equation}
g_t = \tanh\left( \tilde{X}_t W_{x_g} + \tilde{H}_{t-1} W_{h_g} + b_g \right).
\label{eq:gt}
\end{equation}
Here, $W_{x*} \in \mathbb{R}^{F \times H}$ and $W_{h*} \in \mathbb{R}^{H \times H}$ are trainable weight matrices for the input and hidden states, respectively, associated with each gate, and $b_* \in \mathbb{R}^{1 \times H}$ denotes the corresponding bias vectors. The activation functions $\sigma(\cdot)$ and $\tanh(\cdot)$ denote the element-wise sigmoid and hyperbolic tangent, respectively. The memory and hidden states are updated using standard LSTM recurrence in (\ref{eq:ct})-(\ref{eq:ht}):
\begin{equation}
c_t = f_t \odot c_{t-1} + i_t \odot g_t
\label{eq:ct}
\end{equation}
\begin{equation}
h_t = o_t \odot \tanh(c_t)
\label{eq:ht}
\end{equation}
where $\odot$ represents element-wise multiplication.

Through this formulation, spatial dependencies are consistently embedded throughout both the input and memory pathways, enabling the recurrent network to model temporally evolving graph dynamics with enhanced structural awareness.

While the DRAMN framework is designed to operate on real-time measurements, its development, training, and evaluation require a diverse and labeled dataset that reflects a wide range of operating conditions and disturbance scenarios. To this end, a comprehensive dataset is generated using time-domain simulations and modal analyses, covering thousands of generation dispatch combinations and fault configurations. The following section outlines the procedure used to construct these datasets.

\section{Dataset Generation}
\subsection{A Ternary Parameterization System}
\label{sec:ternary_grid}
The penetration levels of SG, GFM, and GFL components within an electrical grid can be effectively illustrated using barycentric coordinates on a ternary grid~\cite{bertozzi2024data,BERTOZZI2025planning}. Each grid point in this system corresponds to a distinct convex combination of penetration levels which sums exactly to 100 (\%), representing the total dispatch in the system.

\subsection{Mixed-Generation Power Systems}
Three renewable-penetrated power systems, each comprising energy resources categorized into three distinct types, are used as the basis to simulate data to train and test the proposed framework. These systems are characterized by dynamical behaviors that span a wide range of bandwidths, fast controls in GFM and GFL converters, as well as sensitivity to grid strength, posing a significant technical challenge to grid operators in ensuring stable and reliable operation in the face of varying load profiles and renewable energy availability.

\subsubsection{A Modified 9-Bus Model}
A modified version of the IEEE standard 9-bus system~\cite{bertozzi2024data} is used as a representative benchmark for assessing the proposed stability prediction framework. In this model, two of the synchronous generator units are each replaced with a GFM and a GFL inverter. In this model, the SG is equipped with a governor, excitation control, and a power system stabilizer. The GFM inverter is droop-control-based, and the GFL includes DC-side dynamics.


\subsubsection{HVDC Model} \label{HVDC_section}
\begin{figure*}[t!]
    \centering
    \includegraphics[width=0.8\linewidth]{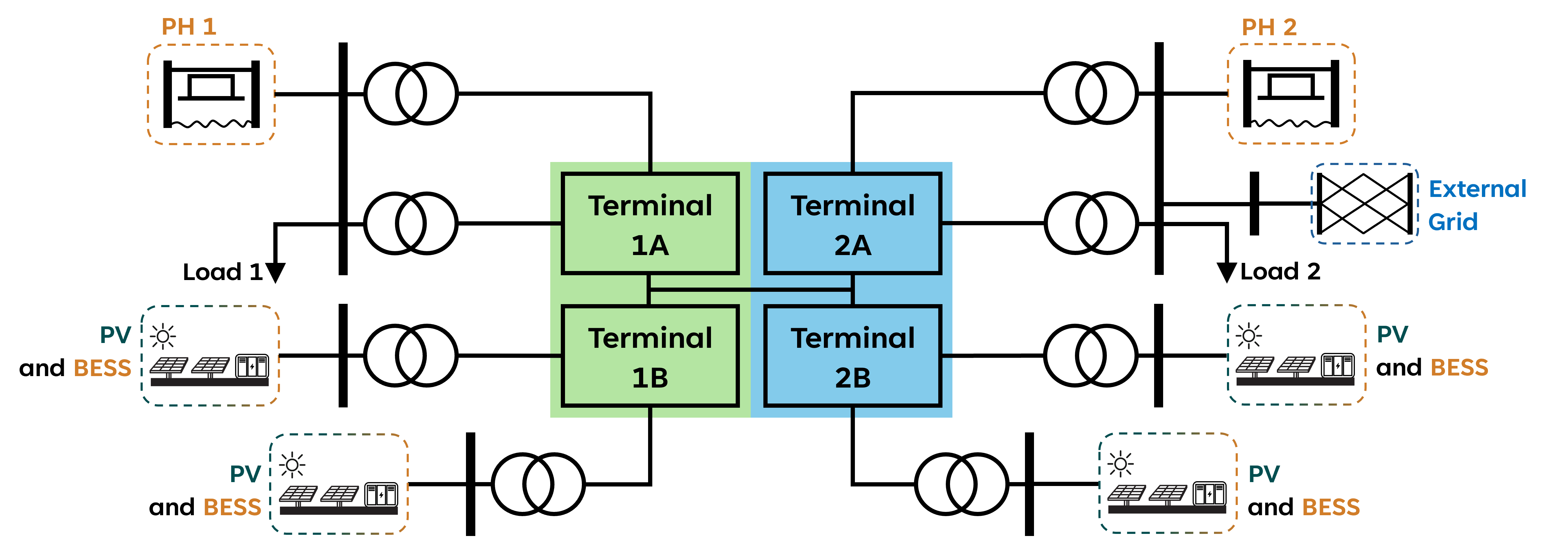}
    \caption{U-shaped HVDC network interconnecting a diverse generation mix composition. The bipolar links 1 and 2 are interconnected by a tie-line. Terminals B of both links are fed by PV generation and BESS. Terminal A of link 1 is connected to a PH station, while terminal A of link 2 is connected to a PH station and to a grid equivalent through a transmission line.}
    \label{fig:HVDC}
\end{figure*}

The U-shaped multi-terminal HVDC system represents a more complex mixed-generation grid comprising GFL renewable energy resources, BESS, SG-based hydro storage, and a strong grid interconnection distributed across four different AC areas interconnected through a four-terminal HVDC network, as shown in Fig.~\ref{fig:HVDC}~\cite{bertozzi2024data}. The system comprises two SGs equipped with the same controllers as in the modified 9-bus model, representing pumped hydro (PH) units. The GFM battery storage units are placed in the same areas as the GFL PV generation, which are modeled as in the modified 9-bus model.

Both the 9-bus and HVDC systems are used to simulate two categories of events, where all simulations are performed for 60 seconds. The first event is a global load step change at time $t = 20.00$ s where the active power demand of all loads in the systems are increased by 10\%. This change is maintained until the end of the simulation. The second is a short-circuit event which occurs at Line 4-5 in the 9-bus system and at the cable connecting the external grid in the HVDC system, occurring at time $t = 20.00$ s and cleared at $t = 20.05$ s. These events are chosen to represent small-signal and transient disturbances, allowing for a more comprehensive understanding of the dynamic behavior and stability response of the system under typical operational stress conditions.


\subsubsection{A Modified 39-Bus System} \label{39-bus intro}
A modified IEEE 39-bus system~\cite{athay1979} is also included to assess scalability. Generators 3, 8, and 9 are replaced by GFM units, and generators 4, 7, and 10 are replaced by GFL units. The dataset generation follows the same procedure described above, with each operating point simulated for 60 s. Unlike the 9-bus and HVDC systems, only unperturbed operating scenarios are simulated for the 39-bus system. This design choice is intended to demonstrate the scalability of the proposed framework under larger network sizes, while limiting the dataset complexity to steady-state variations in the generation mix. 

In all models, the voltage and frequency measurements at all buses, as well as the active (P) and reactive (Q) powers and frequencies of all loads, generating units, and storage units are recorded as time-series data for each simulation. The optimal combination of these measurements as input to the predictive framework is subsequently investigated in Section~\ref{sec:results}.

\subsection{Data Labeling and Preparation}
To ensure consistent evaluation across all test systems, a unified parameterization scheme is applied to generate the corresponding datasets. The ternary grid described in Section \ref{sec:ternary_grid} discretizes the participation levels of SG, GFM, and GFL units at 1\% resolution, with a minimum contribution of 1\% per type, such that each combination sums to 100\%. This results in a total of $\binom{99}{2}=4851$ unique generation-mix combinations.

Since the objective of DRAMN is to predict system stability based on current time-series data, all sliding windows extracted from measurements corresponding to a given generation mix scenario are assigned a common stability label. This label is assigned by the modal analysis function of \textsc{PowerFactory}~\cite{powerfactory_manual}, which computes eigenvalues at the end of each simulation. A case is classified as ``1'' (unstable) if it meets any of the following criteria:
\begin{enumerate}
    \item Possesses an eigenvalue with $\mathrm{Re}(\lambda) > 0$
    \item Exhibits voltages $<0.95$\,p.u. or $>1.05$\,p.u.~\cite{ansiC84.1_2020}
    \item Exhibits frequencies $<59.80$\,Hz or $>60.20$\,Hz~\cite{nercPRC024_3}
    \item Exhibits damping ratios $\zeta < 3\%$~\cite{IEEE1110_2019}
\end{enumerate}
\noindent in which case all post-event data windows are labeled accordingly; otherwise, the case is labeled ``0'' (stable).

\begin{figure}[t!]
    \centering
    \includegraphics[width=\linewidth]{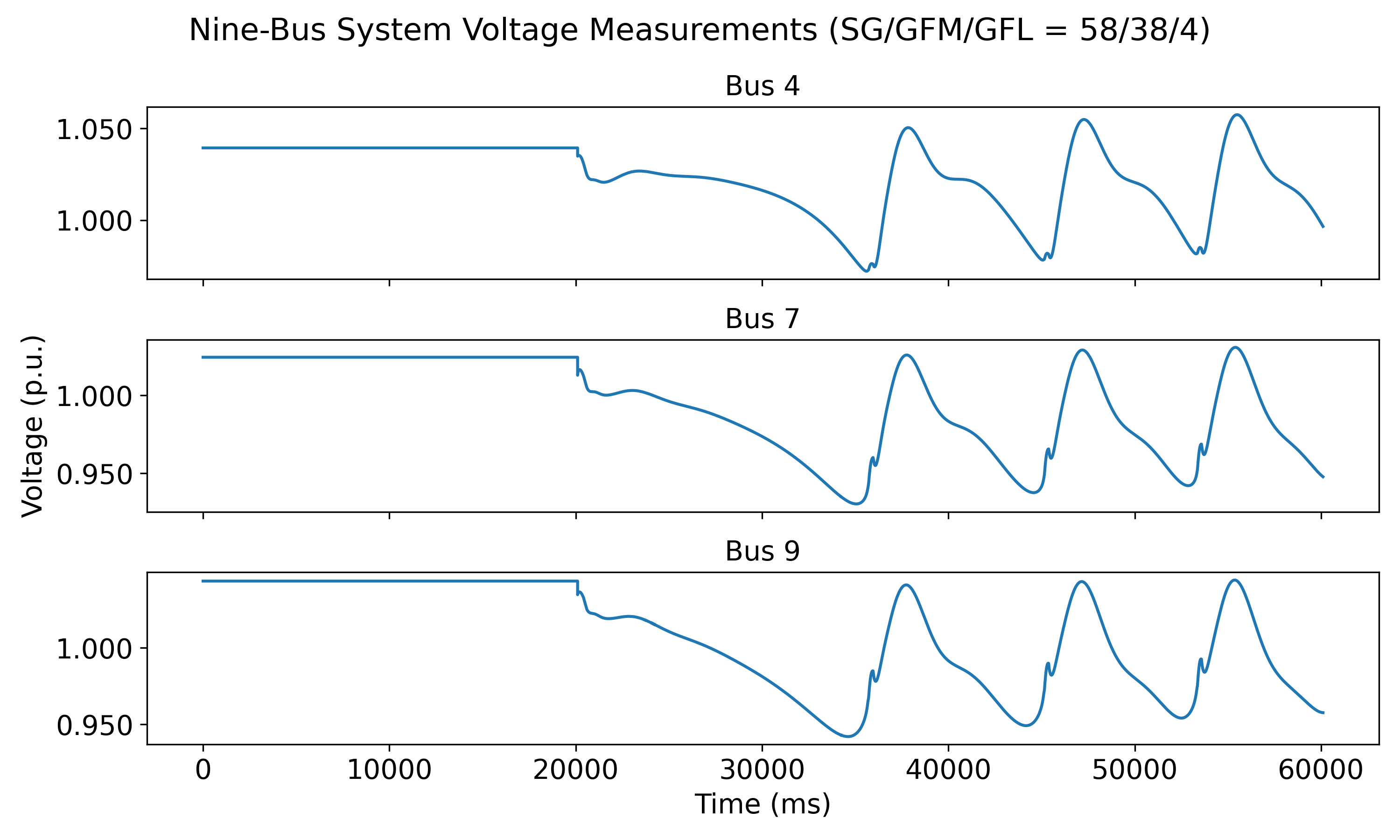}
    \caption{Voltage trajectories measured at Buses 4, 7, and 9 during a load increase event under the generation mix SG/GFM/GFL = 58/38/4.} 
    \label{fig:unstable load increase example}
\end{figure}

As an illustrative example, Fig.~\ref{fig:unstable load increase example} shows voltage measurements recorded at Buses 4, 7, and 9 of the modified 9-bus system during a scenario characterized by a generation mix of SG/GFM/GFL = 58/38/4 with a load increase event at $t = 20000$ ms. Windows of duration $T = 1000$ ms are applied on the data points from $t = 19001$ ms to $t = 30000$ ms at steps of $100$ ms. In this particular case, the largest eigenvalue exhibits a positive real part of 0.3687, leading to all associated time windows from $t = 19001$ ms to $t = 30000$ ms being labeled as ``1.'' Since the window size is $1000$ ms and the event occurs at $t = 20000$ ms, $t = 19001$ ms represents the first millisecond where the event is detected by the window while $t = 30000$ ms represents the first millisecond where the window leaves the time of occurrence. As a means to investigate the framework's ability to generalize its predictions, the data points from $t = 0$ ms to $t = 19000$ ms and $t = 30001$ ms to $t = 60000$ ms are not used to train DRAMN, but are fed into the network for testing, as demonstrated in Section~\ref{sec:results}. These ranges represent the default unperturbed state and the post-event state of the system, respectively.

Each generation mix scenario yields 11 time-series windows. To reduce overfitting and improve efficiency, only 1 in 20 load increase and short-circuit cases is selected at random. After removing non-convergent cases, the modified 9-bus dataset contains 44,523 samples, the HVDC dataset comprises 50,668 samples, while the 39-bus dataset has 13004 samples. All system events are simulated for each generation mix permutation in both systems. The resulting ``stable'' and ``unstable'' cases are summarized in Table~\ref{tab:distributions}.
\begin{table}[!t]
    \renewcommand{\arraystretch}{1.3}
    \caption{Stability Distribution by Case and Event}\label{tab:distributions}
    \centering
        \begin{tabular}{c c c c}
            \hline\hline \\[-3mm]
            \multicolumn{1}{c}{System} & \multicolumn{1}{c}{Event} & \multicolumn{1}{c}{Stable Cases} & \multicolumn{1}{c}{Unstable/Diverged Cases} \\[1.4ex]\hline
            9-bus & Load increase & $4443$ & $408$ \\
            9-bus & Short-circuit & $4190$ & $661$ \\
            HVDC & Load increase & $401$ & $4450$ \\
            HVDC & Short-circuit & $688$ & $4163$ \\
            39-bus & Unperturbed & $1304$ & $3547$\\
            \hline\hline
        \end{tabular}
\end{table}
From the table, a total of 8.41\% and 13.63\% of the simulated scenarios in the 9-bus system exhibited instability under load increase and short-circuit events, respectively, while 91.73\% and 85.82\% of the HVDC scenarios became unstable under these events.

\section{Results and Discussion} \label{sec:results}
This section presents tests on the renewable-penetrated models, conducted on the Ibex High Performance Computing (HPC) Cluster of KAUST using GPUs with 256 GB RAM and 8 CPU cores. All models are trained with AdamW using a learning rate of $1\times10^{-3}$ with MAE as the loss function. Training runs for 100 epochs with a batch size of 32 with early stopping enabled. The dataset is split 80/20 into training and testing sets, and 10\% of the training portion is reserved for validation. Input time series are standardized via z-score normalization. These settings are selected prior to tuning experiments to balance convergence and generalization. 

\subsection{Input Window Size Selection}
The input window size is a critical parameter affecting both the capture of relevant system dynamics and computational efficiency. Short windows may fail to represent oscillatory patterns and transient behaviors, while long ones introduce redundant information and impede feature extraction. To investigate the optimal window size, DRAMN is trained using simulation data from all systems with sliding-window sizes $T = 100, 200, ..., 2000$ ms. Selected results for $T = 100, 500, 1000,$ and $2000$ ms on the modified 9-bus system are reported in Table~\ref{tab:window_size_selected}, where the highest values of each metric are shown in blue. $T = 1000$ ms is selected as the default window size for all subsequent tests as it achieves the highest F1 score while maintaining near-optimal values across all other metrics.
\begin{table}[!t]
    \renewcommand{\arraystretch}{1.3}
    \caption{Window Size Impact on Model Performance (Selected Sizes)}\label{tab:window_size_selected}
    \centering
    \small
    \begin{tabular}{l c c c c}
        \hline\hline \\[-3mm]
        & 100 ms & 500 ms & 1000 ms & 2000 ms \\[1.4ex] \hline
        Accuracy     & $0.9913$ & $0.9907$ & $\textcolor{blue}{0.9985}$ & $0.9975$ \\
        Precision    & $0.9371$ & $0.9281$ & $\textcolor{blue}{0.9928}$ & $0.9899$ \\
        Recall       & $0.9412$ & $0.9342$ & $0.9884$ & $\textcolor{blue}{0.9901}$ \\
        F1 Score     & $0.9392$ & $0.9311$ & $\textcolor{blue}{0.9835}$ & $0.9832$ \\
        Specificity  & $0.9951$ & $0.9948$ & $\textcolor{blue}{0.9994}$ & $0.9974$ \\
        AUROC        & $0.9930$ & $0.9881$ & $0.9995$ & $\textcolor{blue}{0.9997}$ \\
        \hline\hline
    \end{tabular}
\end{table}

\subsection{Model Selection and Benchmarking}
The proposed framework is benchmarked against random forest and gradient boost algorithms, as well as four recent predictive architectures, to comprehensively evaluate its performance, as shown in Table~\ref{tab:comprehensive_model_comparison}, where the highest values of each metric are shown in blue. These models represent a diverse approach to stability assessment.
\begin{table*}[!t]
    \renewcommand{\arraystretch}{1.3}
    \caption{Benchmark Model Performance Comparison}\label{tab:comprehensive_model_comparison}
    \centering
    \small
    \begin{tabular*}{\textwidth}{@{\extracolsep{\fill}} l c c c c c c c c c c @{}}
        \hline\hline \\[-3mm]
        \multicolumn{1}{c}{Metric} & 
        \multicolumn{1}{c}{\begin{tabular}[c]{@{}c@{}}DMD\\Random\\Forest\end{tabular}} & 
        \multicolumn{1}{c}{\begin{tabular}[c]{@{}c@{}}DMD\\Gradient\\Boost\end{tabular}} & 
        \multicolumn{1}{c}{\begin{tabular}[c]{@{}c@{}}ST-\\AGNet\end{tabular}} & 
        \multicolumn{1}{c}{\begin{tabular}[c]{@{}c@{}}CNN-\\GAT\end{tabular}} & 
        \multicolumn{1}{c}{\begin{tabular}[c]{@{}c@{}}Transformer-\\based TSA\end{tabular}} & 
        \multicolumn{1}{c}{PowerGNN} & 
        \multicolumn{1}{c}{\begin{tabular}[c]{@{}c@{}}DRAMN-\\N-DMD\end{tabular}} & 
        \multicolumn{1}{c}{\begin{tabular}[c]{@{}c@{}}DRAMN-\\KE-DMD\end{tabular}} & 
        \multicolumn{1}{c}{\begin{tabular}[c]{@{}c@{}}DRAMN-\\TD-DMD\end{tabular}} &
        \multicolumn{1}{c}{\begin{tabular}[c]{@{}c@{}}DRAMN-\\DMD\end{tabular}} \\[1.4ex]\hline
        Accuracy      & $0.9938$ & $0.9972$ & $0.9965$ & $0.9976$ & $0.9889$ & $0.9965$ & $0.9980$ & $0.9980$ & $0.9978$ & ${\color{blue}0.9985}$ \\
        Precision     & $0.9801$ & $0.9865$ & $0.9880$ & $0.9956$ & $0.9750$ & $0.9890$ & $0.9842$ & $0.9856$ & ${\color{blue}0.9985}$ & $0.9928$ \\
        Recall        & $0.9402$ & $0.9721$ & $0.9467$ & $0.9741$ & $0.9650$ & $0.9720$ & $0.9899$ & $0.9885$ & $0.9726$ & ${\color{blue}0.9944}$ \\
        F1 Score      & $0.9597$ & $0.9792$ & $0.9669$ & ${\color{blue}0.9847}$ & $0.9700$ & $0.9804$ & $0.9871$ & $0.9871$ & $0.9854$ & $0.9835$ \\
        Specificity   & $0.9968$ & $0.9966$ & $0.9989$ & $0.9996$ & $0.9986$ & $0.9989$ & $0.9987$ & $0.9988$ & $0.9989$ & ${\color{blue}0.9994}$ \\
        AUROC         & $0.9979$ & $0.9977$ & $0.9995$ & $0.9994$ & $0.9985$ & ${\color{blue}0.9998}$ & $0.9993$ & $0.9936$ & $0.9863$ & $0.9995$ \\
        \hline\hline
    \end{tabular*}
\end{table*}

ST-AGNet~\cite{zhang2024stagnet} uses spatial-temporal attention and adaptive graph structure learning to capture complex dependencies in power networks. For this study, predefined electrical connections are used to build graph topology, while temporal features are extracted via convolutional layers. The attention mechanism operates across both spatial and temporal dimensions to model dynamic system relationships. The CNN+GAT~\cite{huang2024multi} model integrates a CNN-based encoder with graph attention networks to classify stability and identify critical generators in a multi-task setup. Buses are mapped to nodes and branches to edges, with CNN handling temporal variation and GAT extracting spatial features. The input tensor is reshaped for temporal coherence, and the task is modified for binary classification. The transformer-based TSA model~\cite{zhao2024swin} employs an encoder-decoder architecture with attention to assess transient stability. Transformer blocks include feed-forward networks and normalization layers. Positional encoding captures sequence structure, while attention mechanisms model temporal dependencies across multiple scales. The PowerGNN model~\cite{li2025power} constructs graph representations via DMD and applies a graph over-sampling method to address power-law node distributions. Multiple adjacency layers—based on participation factors, mode characteristics, and eigenvalues—are normalized and used for graph convolution, capturing both topological and dynamic system behavior.

The DMD algorithm is also replaced with three advanced variants to evaluate the optimal method for constructing adjacency matrices, as shown in Table~\ref{tab:comprehensive_model_comparison}. Neural DMD (N-DMD)~\cite{iwata2023neural} integrates neural network-based autoencoders with spectral decomposition, where observations are encoded into a lifted space and forecasts are made through DMD analysis of the encoded data. Kernel extended DMD (KE-DMD)~\cite{williams2015kernel} leverages radial basis function kernels to capture nonlinear dynamics without explicit feature mappings. Time-Delayed DMD (TD-DMD)~\cite{bronstein2022spatiotemporal} embeds temporal delays into the state representation using parameters selected via mutual information analysis, enriching the spectral structure for stability assessment.

The performance comparison in Table~\ref{tab:comprehensive_model_comparison} demonstrates that the proposed DRAMN-DMD model achieves superior performance in accuracy, recall, and specificity compared to the benchmark models. In the case of stability prediction, recall is a particularly crucial metric since it indicates the model's capabilities in avoiding false negatives (undetected destabilizing situations). 

\subsection{Ablation Studies}
To assess the performance impact of each architectural component in the DRAMN framework, the results for DRAMN and its ablated variants are summarized in Table~\ref{tab:ablation_study}, where the highest values of each metric are shown in blue. Various deep learning architectures such as LSTM-RNN, GCN, and GCN-LSTM are evaluated. Different versions of DMD such as the default DMD, extended DMD, and Hankel DMD are also compared. The inclusion of DRAMN with $L_{seq}=1$ isolates the contribution of multi-layer adjacency construction from temporal recurrence, thereby validating that both spatial graph processing and temporal dependencies are essential for accurate stability forecasting.
\begin{table*}[!t]
    \renewcommand{\arraystretch}{1.3}
    \caption{Ablation Study Model Performance Comparison}\label{tab:ablation_study}
    \centering
    \small
    \begin{tabular*}{\textwidth}{@{\extracolsep{\fill}} l c c c c c c c c @{}}
        \hline\hline \\[-3mm]
        \multicolumn{1}{c}{Metric} & 
        \multicolumn{1}{c}{DMD} & 
        \multicolumn{1}{c}{Extended DMD} & 
        \multicolumn{1}{c}{Hankel DMD} & 
        \multicolumn{1}{c}{LSTM} & 
        \multicolumn{1}{c}{GCN} & 
        \multicolumn{1}{c}{GCN-LSTM} & 
        \multicolumn{1}{c}{DRAMN ($L_{seq}=1$)} & 
        \multicolumn{1}{c}{DRAMN ($L_{seq}=5$)} \\[1.4ex]\hline
        Accuracy      & $0.9604$ & $0.9801$ & $0.9709$ & $0.9862$ & $0.9787$ & $0.9931$ & Non-convergent & ${\color{blue}0.9985}$ \\
        Precision     & $0.9150$ & $0.9600$ & $0.9400$ & $0.9454$ & $0.8569$ & $0.9265$ & Non-convergent & ${\color{blue}0.9928}$ \\
        Recall        & $0.8900$ & $0.9400$ & $0.9200$ & $0.8732$ & $0.8718$ & $0.9847$ & Non-convergent & ${\color{blue}0.9944}$ \\
        F1 Score      & $0.9000$ & $0.9500$ & $0.9300$ & $0.9077$ & $0.8641$ & $0.9547$ & Non-convergent & ${\color{blue}0.9835}$ \\
        Specificity   & $0.9949$ & $0.9938$ & $0.9950$ & $0.9957$ & $0.9877$ & $0.9938$ & Non-convergent & ${\color{blue}0.9994}$ \\
        AUROC         & $0.9632$ & $0.9944$ & $0.9861$ & $0.9900$ & $0.9850$ & $0.9950$ & Non-convergent & ${\color{blue}0.9995}$ \\
        \hline\hline
    \end{tabular*}
\end{table*}

The proposed DRAMN outperforms all baselines across all metrics with $L_{seq}=5$, while the variant with $L_{seq}=1$ fails to converge, demonstrating that temporal recurrence is essential for model stability. These results affirm the effectiveness of both DMD-based multi-layer adjacency construction and temporal graph memory structure for robust and accurate stability prediction.

\subsection{Case Study: HVDC System}
\label{sec:HVDC}
The DRAMN framework is applied on data from the HVDC system detailed in Section~\ref{HVDC_section} to validate its applicability on real-world power systems. A total of 72 features are recorded from each simulation, including the active and reactive powers of generating and storage units and loads, as well as the voltage and frequency of AC and DC terminals and cables. The framework is first trained and tested with short-circuit and load change event datasets separately to validate its ability to predict transient and small-signal stability, as shown in the first two columns of Table~\ref{tab:hvdc_performance}, where the highest values of each metric (excluding the isolated short-circuit and load increase results) are highlighted in green. These datasets are then combined and fed into the framework. 
\begin{table*}[!t]
    \renewcommand{\arraystretch}{1.3}
    \caption{HVDC System Model Performance by Number of Nodes and Event}\label{tab:hvdc_performance}
    \centering
    \small
    \begin{tabular*}{\textwidth}{@{\extracolsep{\fill}} l c c c c c c c c c @{}}
        \hline\hline \\[-3mm]
        \multicolumn{1}{c}{Metric} & 
        \multicolumn{1}{c}{\begin{tabular}[c]{@{}c@{}}Short-Circuit\\All Nodes\end{tabular}} & 
        \multicolumn{1}{c}{\begin{tabular}[c]{@{}c@{}}Load Increase\\All Nodes\end{tabular}} & 
        \multicolumn{1}{c}{\begin{tabular}[c]{@{}c@{}}Combined\\9 Nodes\end{tabular}} & 
        \multicolumn{1}{c}{\begin{tabular}[c]{@{}c@{}}Combined\\11 Nodes\end{tabular}} & 
        \multicolumn{1}{c}{\begin{tabular}[c]{@{}c@{}}Combined\\13 Nodes\end{tabular}} & 
        \multicolumn{1}{c}{\begin{tabular}[c]{@{}c@{}}Combined\\15 Nodes\end{tabular}} & 
        \multicolumn{1}{c}{\begin{tabular}[c]{@{}c@{}}Combined\\17 Nodes\end{tabular}} & 
        \multicolumn{1}{c}{\begin{tabular}[c]{@{}c@{}}Combined\\19 Nodes\end{tabular}} & 
        \multicolumn{1}{c}{\begin{tabular}[c]{@{}c@{}}Combined\\All Nodes\end{tabular}} \\[1.4ex]\hline
        Accuracy & $0.9957$ & $0.9977$ & $0.9839$ & $0.9905$ & ${\color{green!50!black}0.9969}$ & $0.9968$ & $0.9966$ & $0.9964$ & $0.9960$ \\
        Precision & $0.9999$ & $0.9996$ & $0.9920$ & $0.9932$ & $0.9941$ & $0.9944$ & $0.9944$ & ${\color{green!50!black}0.9947}$ & $0.9944$ \\
        Recall & $0.9945$ & $0.9973$ & $0.9654$ & $0.9819$ & ${\color{green!50!black}0.9979}$ & $0.9974$ & $0.9969$ & $0.9964$ & $0.9951$ \\
        F1 Score & $0.9972$ & $0.9985$ & $0.9785$ & $0.9875$ & ${\color{green!50!black}0.9960}$ & $0.9959$ & $0.9956$ & $0.9954$ & $0.9947$ \\
        Specificity & $0.9996$ & $0.9988$ & $0.9952$ & $0.9958$ & $0.9963$ & $0.9965$ & $0.9965$ & $0.9965$ & ${\color{green!50!black}0.9966}$ \\
        AUROC & $0.9971$ & $0.9980$ & $0.9803$ & $0.9889$ & ${\color{green!50!black}0.9971}$ & $0.9970$ & $0.9967$ & $0.9964$ & $0.9958$ \\
        \hline\hline
    \end{tabular*}
\end{table*}

While comprehensive measurement from all system components is feasible in simulation, real-world PMU deployment faces economic and operational constraints chiefly due to the cost of communications infrastructure in areas lacking sufficient networking. For the purpose of enabling accurate stability predictions, DMD is explored as a data-driven alternative for dominant feature extraction, leveraging its capabilities in modal analysis, signal reconstruction, and harmonic identification to extract dynamic insights from measurements without requiring an analytical system model. In the proposed graph-based framework, each of the 72 features is treated as a node in a time-varying network, with node strength defined as the sum of incident edge weights. This enables identification of critical measurement locations based on dynamic behavior rather than static topology.

Given the five-layer structure of each adjacency matrix, five corresponding graph networks are instantiated per time step to capture multi-dimensional connectivity. The cumulative node strength $W_n$ for the $n^{th}$ node of each graph in each of the five layers is defined as $W_n = \sum_{t=19000}^{30000} \sum_{j=1}^{N} A^{(l)}_{nj}(t)$,  where $A_{nj}^{(l)}(t)$ represents the edge weight between nodes $n$ and $j$ in layer $l$ at time $t$, and $N$ denotes the total number of nodes in the network. To ensure scale-invariant node selection, the raw cumulative node strengths in each layer are normalized using min-max normalization, such that $\tilde{W}_n = \frac{W_n - \min(W)}{\max(W) - \min(W)}$, where $\tilde{W}_n$ represents the normalized strength of node $n$. These node strengths are then summed to produce an aggregated connectivity graph, defined as $\tilde{W}n^{(composite)} = \sum_{l=1}^{5} \tilde{W}_n^{(l)}$. 

To identify dominant nodes for small-signal and transient stability analysis, aggregated graphs are constructed separately for load increase and short-circuit datasets. Analysis shows that 15 of the top 20 nodes overlap, indicating strong correlation. Thus, the top nodes from the combined dataset are used as DRAMN input features.

\begin{figure}[h!] 
    \centering
    \includegraphics[width=0.4\textwidth]{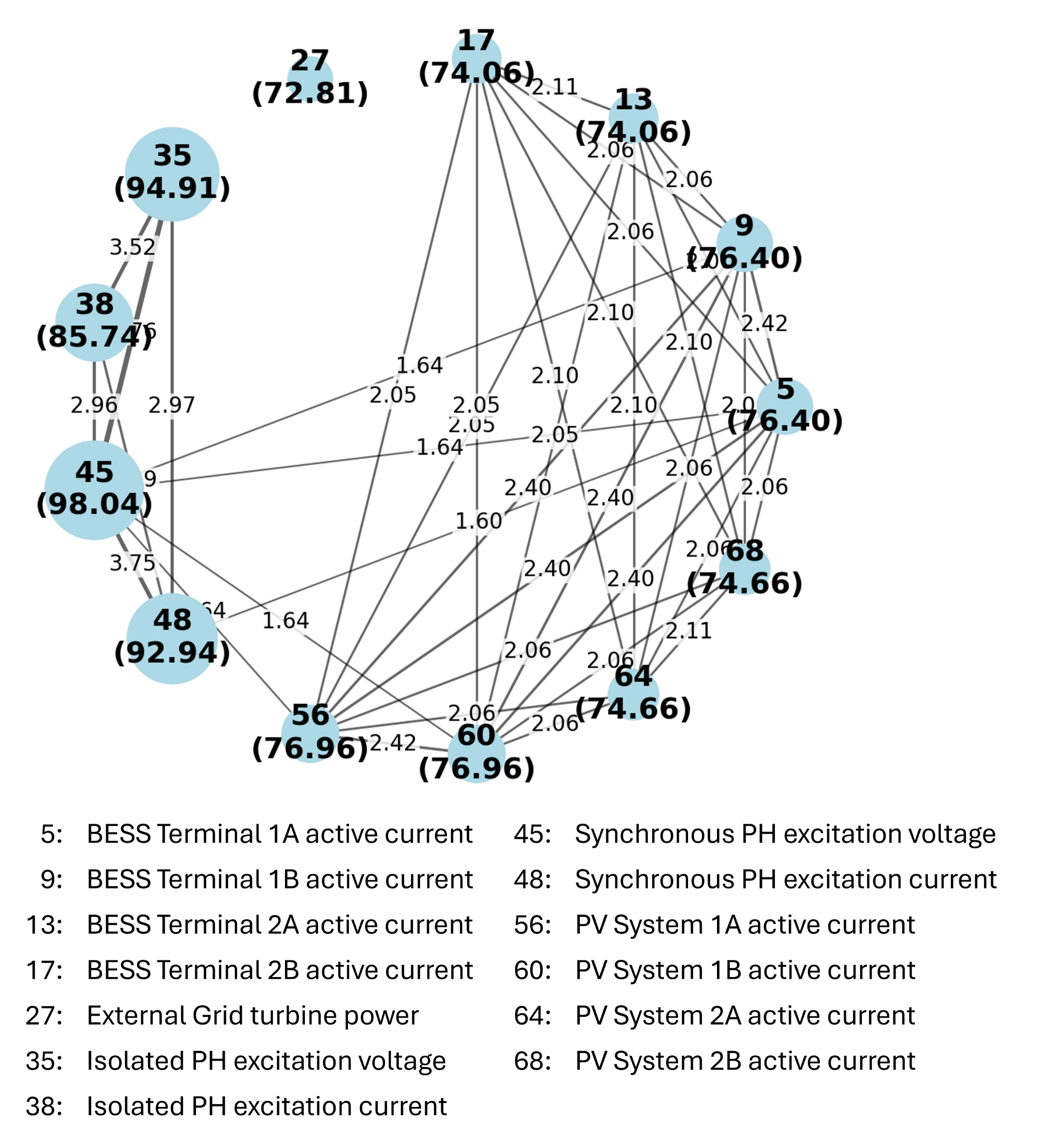}
    \caption{Top 13 dominant nodes of the HVDC system.}
    \label{fig:ooo}
\end{figure}

Columns 3 to 9 of Table~\ref{tab:hvdc_performance} report the framework performance on the combined test data using measurements from the top 9 to all 72 nodes. The highest metric values among these seven columns are highlighted in green. Performance is significantly degraded when using only 9 or 11 nodes, while configurations with 13, 15, 17, 19, and all nodes yield comparable results. The 13-node configuration achieves the highest accuracy, recall, F1 score, and AUROC, and is thus selected as the optimal input feature combination. This reduces the input dimensionality from 72 to 13 features. The corresponding aggregated graph is illustrated in Fig.~\ref{fig:ooo}, with only the top 50\% of edges displayed for clarity.

\subsubsection{Noise Sensitivity Analysis}
\begin{figure}[h!]
    \centering
    \includegraphics[width=0.35\textwidth]{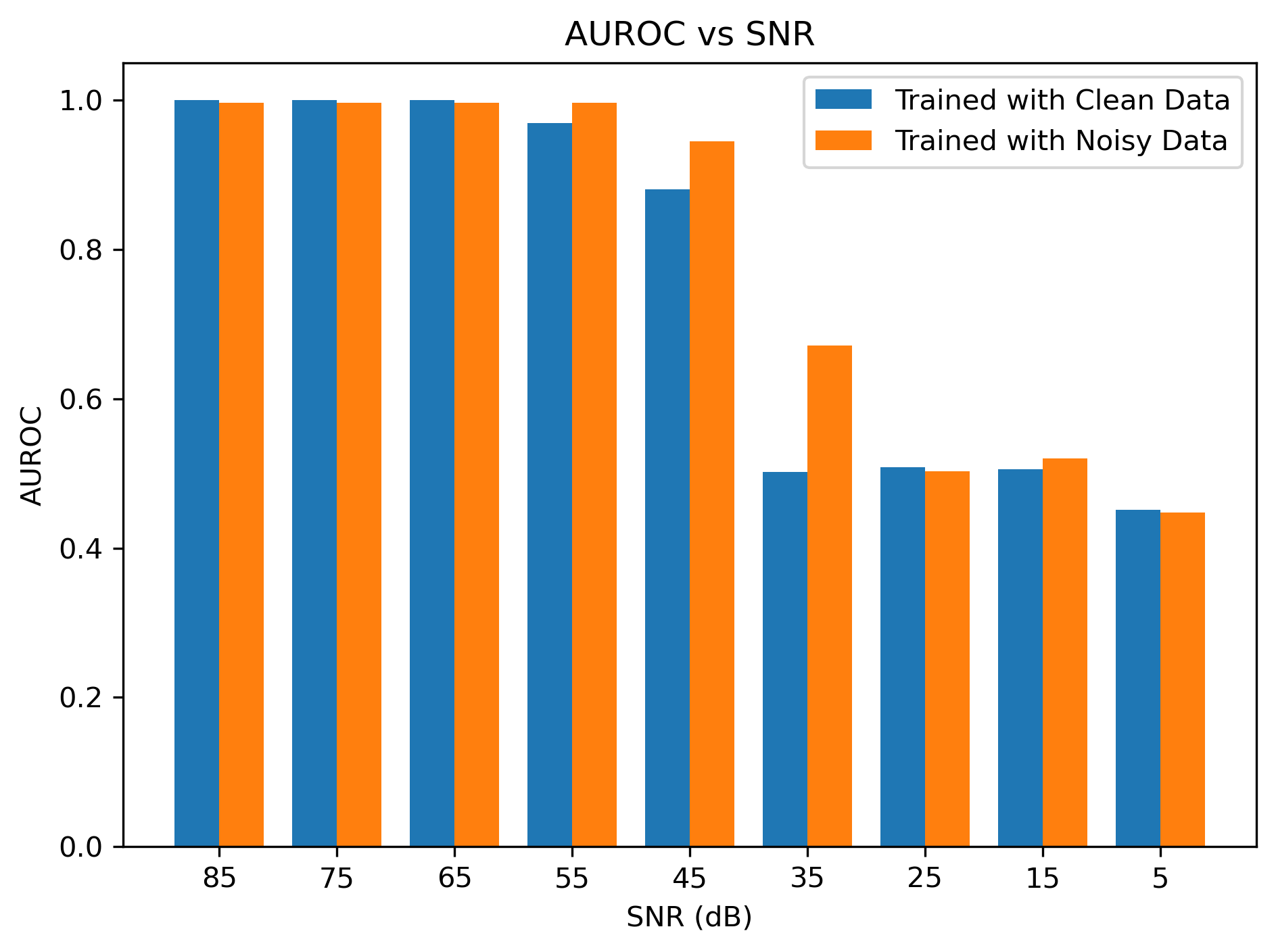}
    \caption{AUROC of DRAMN trained on clean versus noise-augmented data, evaluated on HVDC measurements corrupted with AWGN at varying SNR levels.}
    \label{fig:AUROC_vs_SNR}
\end{figure}

To assess robustness, additive white Gaussian noise is injected into the HVDC measurements at \(\mathrm{SNR}\in\{5,15,\dots,85\}\,\mathrm{dB}\). The AUROC for DRAMN trained on clean data versus noise-augmented data and evaluated on noisy inputs is summarized in Fig.~\ref{fig:AUROC_vs_SNR}.

For \(\mathrm{SNR}\geq 55\,\mathrm{dB}\), both models yield comparable AUROC. As \(\mathrm{SNR}\) decreases below \(55\,\mathrm{dB}\), AUROC degrades, with the clean-trained model deteriorating more rapidly than its noise-trained counterpart. At \(25\,\mathrm{dB}\), both approaches converge to \(\mathrm{AUROC}\approx 0.5\), indicating a loss of discriminative capability for identifying unstable scenarios.

\subsubsection{Computational Efficiency}
The computational performance of the proposed DRAMN framework was evaluated on the Ibex HPC cluster to assess real-time deployment feasibility. All reported timing measurements represent the mean computational time averaged over 10,000 independent test samples on the same machine, ensuring statistical reliability of the performance estimates.
\begin{table}[!t]
    \renewcommand{\arraystretch}{1.3}
    \caption{Computational Runtime Performance}
    \label{tab:runtimes}
    \centering
    \begin{tabular}{p{2.5cm} c c c}
        \hline\hline \\[-3mm]
        \multicolumn{1}{c}{\textbf{Runtime (ms)}} & \multicolumn{1}{c}{\textbf{9-Bus}} & \multicolumn{1}{c}{\textbf{HVDC (72)}} & \multicolumn{1}{c}{\textbf{HVDC (13)}} \\[1.4ex]\hline
        Adjacency Matrix & 3.65 & 7.13 & 4.77 \\
        Prediction & 0.52 & 1.82 & 0.97 \\
        Total & 4.17 & 8.95 & 5.74 \\
        \hline\hline
    \end{tabular}
\end{table}

Adjacency matrix construction via sliding-window DMD requires 3.65 ms, 7.13 ms, and 4.77 ms per 1-second window for the 9-bus system and HVDC systems with 72 and 13 nodes as input, respectively, while the trained DRAMN model achieves stability prediction in 0.52 ms, 1.82 ms, and 0.97 ms. This shows that the identification of dominant nodes significantly reduces overall runtime.

\begin{figure}[h!]
    \centering
    \includegraphics[width=0.35\textwidth]{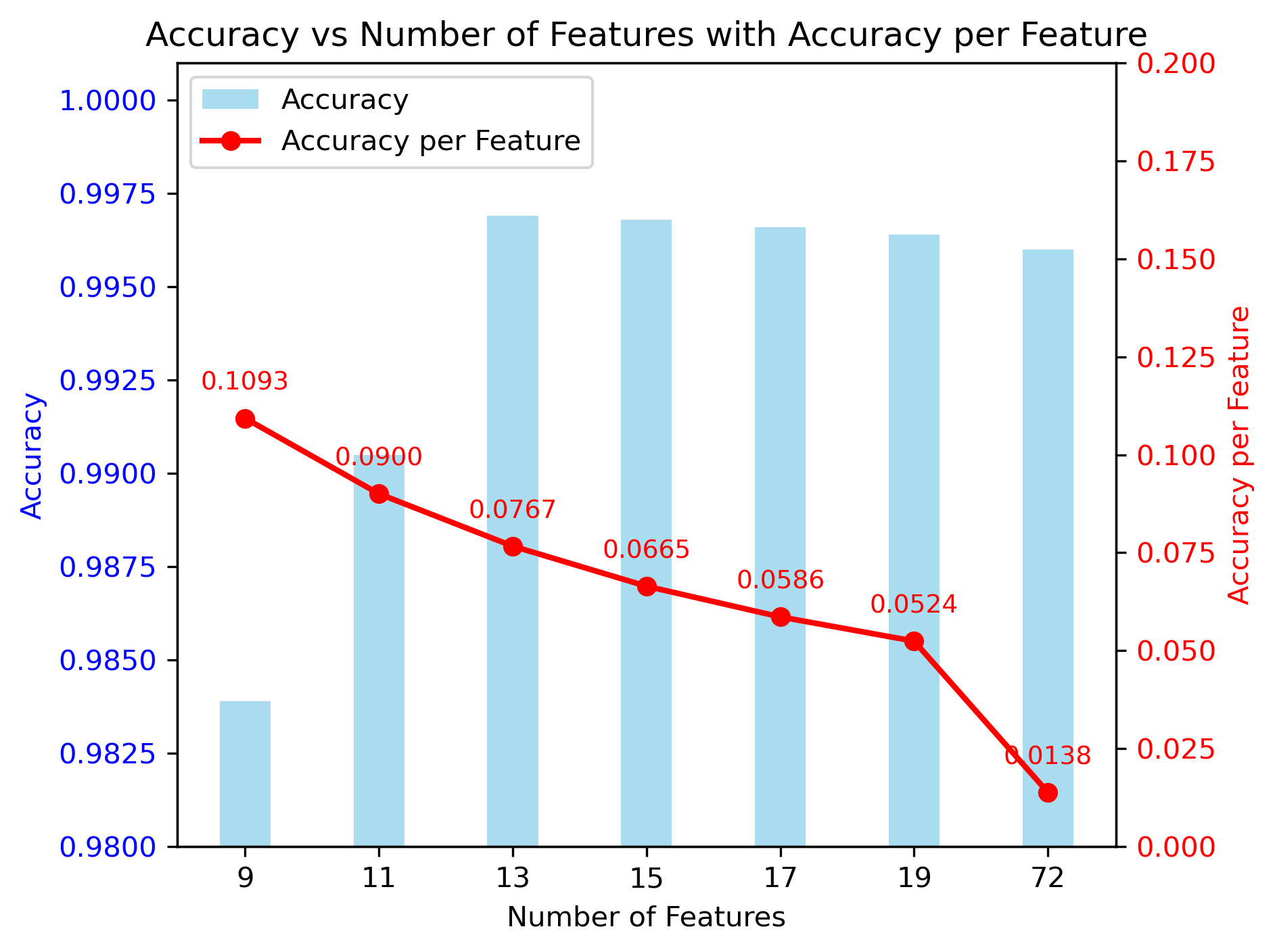}
    \caption{Accuracy per feature of DRAMN trained on different number of input features.}
    \label{fig:accuracy_per_feature}
\end{figure}

In addition to runtime efficiency, the feature selection analysis quantifies the predictive utility of each input. Fig.~\ref{fig:accuracy_per_feature} shows the average per‑feature contribution to DRAMN accuracy increases as the input set is reduced, with the 13‑feature configuration yielding high per-feature value while providing the highest overall accuracy.

\subsection{Case Study: 39-Bus System}
To evaluate scalability, the framework is further tested on the modified IEEE 39-bus system introduced in Section~\ref{39-bus intro}. In this case, only unperturbed operating scenarios are simulated to highlight the scalability of the framework with respect to system size rather than disturbance complexity. Each operating point on the ternary grid is simulated for 60~s, and voltage measurements from all 39 buses are used as inputs.

The framework achieves an AUROC of 0.998 when trained on the full set of bus voltages, demonstrating that the proposed architecture generalizes effectively to larger networks despite increased dimensionality. Furthermore, when the dominant 19 buses are identified through the feature reduction procedure described in Section~\ref{sec:HVDC}, the framework still attains an AUROC of 0.997, confirming that high predictive accuracy can be maintained with significantly fewer measurement points. 

These results indicate that DRAMN preserves both accuracy and feature efficiency as the system size grows, providing evidence of its scalability to larger grids. While disturbance scenarios are not included for the 39-bus system in this study, the scalability assessment complements the fault-driven evaluations on the 9-bus and HVDC systems, together offering a comprehensive view of model robustness.

\subsection{Discussion and Limitations}
While the proposed DRAMN framework demonstrates strong predictive performance and generalization across diverse scenarios, several limitations warrant further discussion:

\subsubsection{Scalability to Larger Systems}  
The current implementation has been validated on small and medium-sized systems. Scaling to large-scale grids may introduce computational challenges due to the size of adjacency matrices and the increased depth of GCN-LSTM layers. Future work could investigate sparse graph encoding, hierarchical aggregation, or localized neighborhood sampling to reduce computational cost on larger systems.

\subsubsection{Feasibility for Real-Time Deployment}  
While the DRAMN architecture is designed with streaming data in mind, real-time deployment requires latency constraints to be met at each stage from DMD-based adjacency construction to forward pass inference. Preliminary timing tests indicate that inference is feasible within operational timescales for medium-sized systems, but online deployment in large networks would require optimized DMD implementations and model quantization or pruning strategies.

\subsubsection{Zero-Shot Topology Changes}  
Preliminary study has been conducted on the sensitivity of DRAMN to topology changes in the system. The cable connecting the external grid to the rest of the system is disconnected rather than restored after the 50 ms short-circuit event. This setup is simulated for all generation mix combinations in steps of 2\%, resulting in 2425 cases. The measurements from this dataset are used directly as test data for DRAMN, resulting in a zero-shot topology-change accuracy of 82.03\%. The decrease in performance is significant in comparison to load-increase and short-circuit scenarios, which stipulates more in-depth study and inclusion of topology-change data in the training dataset. 

These limitations highlight avenues for future research and system-specific adaptation. Nonetheless, the framework provides a promising foundation for interpretable, physics-informed stability forecasting in evolving power systems.

\section{Conclusion}
This paper introduces DRAMN, a hybrid graph-recurrent forecasting model that integrates dynamic mode decomposition with spatio-temporal learning for stability prediction in mixed-generation power systems. By embedding physically meaningful modal features into evolving adjacency matrices, DRAMN bridges the gap between interpretability and predictive performance. The framework generalizes across diverse generation mixes and disturbance scenarios, demonstrating its suitability for both planning and real-time operational contexts. This work also highlights the value of physics-informed learning in modern grid applications. As inverter-based resources continue to reshape power system dynamics, physics-based tools such as DRAMN provide a scalable pathway toward adaptive, data-driven stability monitoring. Future work will focus on scaling to large systems, improving robustness to noise and data gaps, and enabling edge-based deployment for real-time grid resilience.
 


\bibliographystyle{IEEEtran}
\bibliography{main}

@inproceedings{bertozzi2024data,
  title={A Data-Driven Advisory Tool for Mixed-Generation Power Systems},
  author={Bertozzi, Otavio and others},
  booktitle={2024 IEEE Energy Conversion Congress and Exposition (ECCE)},
  pages={1695--1701},
  year={2024},
  organization={IEEE}
}

@article{BERTOZZI2025planning,
title = {Data-driven planning of mixed-generation power systems: Towards 100\% RES-based grids},
journal = {International Journal of Electrical Power \& Energy Systems},
volume = {172},
pages = {111207},
year = {2025},
issn = {0142-0615},
doi = {https://doi.org/10.1016/j.ijepes.2025.111207},
url = {https://www.sciencedirect.com/science/article/pii/S0142061525007550},
author = {Otavio Bertozzi and others}
}

@article{huang2024multi,
  title={A multi-task transient stability assessment method adapted to renewable energy integration},
  author={Huang, L. and Wang, J. and Zhang, Y. and Li, X. and Zhao, J.},
  journal={Frontiers in Energy Research},
  volume={11},
  pages={1321998},
  year={2024},
  publisher={Frontiers},
  doi={10.3389/fenrg.2023.1321998}
}

@article{zhang2024stagnet,
  title={ST-AGNet: Dynamic power system state prediction with spatial–temporal attention graph-based network},
  author={Zhang, Shiyao and Zhang, Shuyu and Yu, James J.Q. and Wei, Xuetao},
  journal={Applied Energy},
  volume={365},
  pages={123252},
  year={2024},
  publisher={Elsevier},
  doi={10.1016/j.apenergy.2024.123252}
}

@article{li2025power,
  title={Power-GNN: a graph over-sampling method to mitigate power-law distribution in graph neural networks},
  author={Li, P and others},
  journal={Applied Intelligence},
  volume={55},
  number={7},
  pages={1--6},
  year={2025},
  month={May}
}

@inproceedings{zhao2024swin,
  title={Swin Transformer Architecture-Based Power System Transient Stability Assessment},
  author={Zhao, Z and others},
  booktitle={2024 IEEE PES 16th Asia-Pacific Power and Energy Engineering Conference (APPEEC)},
  pages={1--5},
  year={2024},
  organization={IEEE}
}

@article{iwata2023neural,
  title={Neural dynamic mode decomposition for end-to-end modeling of nonlinear dynamics},
  author={Iwata, Tomoharu and Kawahara, Yoshinobu},
  journal={Journal of Computational Dynamics},
  volume={10},
  number={2},
  pages={247--265},
  year={2023},
  publisher={American Institute of Mathematical Sciences},
  doi={10.3934/jcd.2022029}
}

@article{williams2015kernel,
  title={A kernel-based method for data-driven {K}oopman spectral analysis},
  author={Williams, Matthew O. and Rowley, Clarence W. and Kevrekidis, Ioannis G.},
  journal={Journal of Computational Dynamics},
  volume={2},
  number={2},
  pages={247--265},
  year={2015},
  publisher={American Institute of Mathematical Sciences},
  doi={10.3934/jcd.2015005}
}

@article{bronstein2022spatiotemporal,
  title={The spatiotemporal coupling in delay-coordinates dynamic mode decomposition},
  author={Bronstein, Emil and others},
  journal={Chaos: An Interdisciplinary Journal of Nonlinear Science},
  volume={33},
  number={1},
  pages={013101},
  year={2023},
  publisher={AIP Publishing},
  doi={10.1063/5.0123101}
}

@standard{ansiC84.1_2020,
  title = {American National Standard for Electric Power Systems and Equipment—Voltage Ratings (60 {Hertz})},
  author = {{American National Standards Institute}},
  organization = {American National Standards Institute},
  year = {2020},
  type = {Standard},
  number = {ANSI C84.1-2020},
  address = {Washington, DC, USA},
  url = {https://webstore.ansi.org/standards/nema/ansic842020},
}

@techreport{nercPRC024_3,
  title = {{NERC} Reliability Standard {PRC-024-3}: Generator Frequency and Voltage Protective Relay Settings},
  author = {{North American Electric Reliability Corporation}},
  institution = {North American Electric Reliability Corporation},
  year = {2023},
  type = {Standard},
  url = {https://www.nerc.com/pa/Stand/Reliability%20Standards/PRC-024-3.pdf},
}

@standard{IEEE1110_2019,
  title = {{IEEE} Guide for Synchronous Generator Modeling Practices and Applications in Power System Stability Analyses},
  author = {{Institute of Electrical and Electronics Engineers}},
  organization = {IEEE},
  year = {2019},
  type = {Standard},
  number = {IEEE Std 1110-2019},
  doi = {10.1109/IEEESTD.2019.8854460},
}

@article{alimi2020review,
title={A Review of Machine Learning Approaches to Power System Security and Stability},
author={Alimi, Oyeniyi and others},
journal={IEEE Access},
volume={8},
pages={512--531},
year={2020},
publisher={IEEE},
doi={10.1109/ACCESS.2020.3003568}
}

@article{zhou2024efficient,
title={An Efficient Method to Estimate Admittance of Black-boxed Inverter-based Resources for Varying Operating Points},
author={Zhou, Weihua and others},
journal={CSEE Journal of Power and Energy Systems},
volume={10},
number={1},
pages={421--426},
year={2024},
publisher={CSEE},
doi={10.17775/CSEEJPES.2023.07090}
}

@ARTICLE{Hatziargyriou2021,
  author={Hatziargyriou, Nikos and others},
  journal={IEEE Transactions on Power Systems}, 
  title={Definition and Classification of Power System Stability – Revisited \& Extended}, 
  year={2021},
  volume={36},
  number={4},
  pages={3271-3281},
  doi={10.1109/TPWRS.2020.3041774}}

@INPROCEEDINGS{Milano2018,
  author={Milano, Federico and others},
  booktitle={2018 Power Systems Computation Conference (PSCC)}, 
  title={Foundations and Challenges of Low-Inertia Systems (Invited Paper)}, 
  year={2018},
  volume={},
  number={},
  pages={1-25},
  doi={10.23919/PSCC.2018.8450880}}

@manual{powerfactory_manual,
  author       = {DIgSILENT},
  title        = {DIgSILENT PowerFactory User Manual},
  year         = {2023},
  organization = {DIgSILENT GmbH},
}

@article{kelada2025revisiting,
  title={Revisiting Small-Signal Modeling for Analyzing Fast Dynamic Interactions in Converter-Dominated Power Systems},
  author={Kelada, Fadi and others},
  journal={IEEE Transactions on Power Systems},
  year={2025},
  publisher={IEEE}
}

@article{lara2023revisiting,
  title={Revisiting power systems time-domain simulation methods and models},
  author={Lara, Jose Daniel and Henriquez-Auba, Rodrigo and Ramasubramanian, Deepak and Dhople, Sairaj and Callaway, Duncan S and Sanders, Seth},
  journal={IEEE Transactions on Power systems},
  volume={39},
  number={2},
  pages={2421--2437},
  year={2023},
  publisher={IEEE}
}

@article{alimi2021review,
  title={A review of research works on supervised learning algorithms for SCADA intrusion detection and classification},
  author={Alimi, Oyeniyi Akeem and others},
  journal={Sustainability},
  volume={13},
  number={17},
  pages={9597},
  year={2021},
  publisher={MDPI}
}

@article{chen2024real,
  title={Real-time Multi-stability Risk Assessment and Visualization of Power Systems: A Graph Neural Network-based Method},
  author={Chen, Qifan and others},
  journal={IEEE Transactions on Power Systems},
  year={2024},
  publisher={IEEE}
}

@article{linaro2023continuous,
  title={Continuous estimation of power system inertia using convolutional neural networks},
  author={Linaro, Daniele and others},
  journal={Nature Communications},
  volume={14},
  number={1},
  pages={4440},
  year={2023},
  publisher={Nature Publishing Group UK London}
}

@article{cao2023physics,
  title={Physics-informed graphical representation-enabled deep reinforcement learning for robust distribution system voltage control},
  author={Cao, Di and others},
  journal={IEEE Transactions on Smart Grid},
  volume={15},
  number={1},
  pages={233--246},
  year={2023},
  publisher={IEEE}
}

@article{falas2025robust,
  title={Robust Power System State Estimation Using Physics-Informed Neural Networks},
  author={Falas, Solon and Asprou, Markos and Konstantinou, Charalambos and Michael, Maria K},
  journal={IEEE Transactions on Industrial Informatics},
  year={2025},
  publisher={IEEE}
}

@article{korda2018convergence,
  title={On convergence of extended dynamic mode decomposition to the Koopman operator},
  author={Korda, Milan and Mezi{\'c}, Igor},
  journal={Journal of Nonlinear Science},
  volume={28},
  number={2},
  pages={687--710},
  year={2018},
  publisher={Springer}
}

@ARTICLE{athay1979,
  author  = {T. Athay and R. Podmore and S. Virmani},
  journal = {IEEE Transactions on Power Apparatus and Systems},
  title   = {A Practical Method for the Direct Analysis of Transient Stability},
  year    = {1979},
  volume  = {PAS-98},
  number  = {2},
  pages   = {573--584},
  doi     = {10.1109/TPAS.1979.319407}
}

@article{nedzhibov2025blind,
  title={Blind Source Separation Using Time-Delayed Dynamic Mode Decomposition.},
  author={Nedzhibov, Gyurhan},
  journal={Computation},
  volume={13},
  number={2},
  year={2025}
}

@article{philipp2025error,
  title={Error analysis of kernel EDMD for prediction and control in the Koopman framework},
  author={Philipp, Friedrich M and others},
  journal={Journal of Nonlinear Science},
  volume={35},
  number={5},
  pages={1--36},
  year={2025},
  publisher={Springer}
}

@article{sakib2025learning,
  title={Learning noise-robust stable koopman operator for control with hankel dmd},
  author={Sakib, Shahriar Akbar and Pan, Shaowu},
  journal={IEEE Transactions on Control Systems Technology},
  year={2025},
  publisher={IEEE}
}

@article{zhu2024scaling,
  title={Scaling graph neural networks for large-scale power systems analysis: empirical laws for emergent abilities},
  author={Zhu, Yuhong and others},
  journal={IEEE Transactions on Power Systems},
  year={2024},
  publisher={IEEE}
}

\newpage

 




\vfill

\end{document}